# Good-Bye Original Sin, Hello Risk On-Off, Financial Fragility, and Crises?

J. Aizenman (USC and NBER), Y. Jinjarak (VUW), D. Park (ADB), and H. Zheng (NUS)[*]

March 2021

## Abstract

We analyze the sovereign bond issuance data of eight major emerging markets (EMs) - Brazil, China, India, Indonesia, Mexico, Russia, South Africa and Turkey from 1970 to 2018. Our analysis suggests that (i) EM local currency bonds tend to be smaller in size, shorter in maturity, or lower in coupon rate than foreign currency bonds; (ii) EMs are more likely to issue local-currency sovereign bonds if their currencies appreciated before the global financial crisis of 2008 (GFC); (iii) inflation-targeting monetary policy increases the likelihood of issuing local-currency debt before GFC but not after; and (iv) EMs that offer higher sovereign yields are more likely to issue local-currency bonds after GFC. Future data will allow us to test and identify structural changes associated with the COVID-19 pandemic and its aftermath.

<u>JEL codes</u>  F21, F31
<u>Keywords</u> Original sin, emerging market, sovereign bond, local-currency bond, financial crisis

[*] Joshua Aizenman Economics and SIR USC University Park Los Angeles, CA 90089-0043 and NBER. aizenman@usc.edu

Yothin Jinjarak School of Economics and Finance Victoria University of Wellington, PO Box 600 23 Lambton Quay, Wellington New Zealand yothin.jinjarak@vuw.ac.nz

Donghyun Park Economics and Research Department Asian Development Bank Manila, Philippines dpark@adb.org

Huanhuan Zheng Lee Kuan Yew School of Public Policy National University of Singapore Singapore arwenzh@gmail.com

# 1. Introduction

The financial opening of emerging markets (EM) in the 1990s provided them with greater access to the global financial system. While access to external finance delivered benefits, mostly front-loaded, it led to significant and growing foreign debt, mostly in hard currency. The wave of sudden stop crises that a dozen EMs suffered in the second half of the 1990s revealed the sizable downside risk associated with significant balance sheet exposures and over-borrowing syndrome related to non-pecuniary externalities. These crises are typically linked to the bailouts of systemic private sector players, thereby socializing their losses and exacerbating the public debt-overhang. More broadly, these developments brought to the fore the hard currency borrowing constraints that emerging markets face: they cannot effectively borrow in their local currency abroad, or even long term domestically. Eichengreen and Hausmann (1999) refer to this phenomenon as the 'Original Sin.'

However, a remarkable adjustment of EM took place in the late 1990s and 2000s. They adopted managed exchange rate flexibility, inflation targeting policy, precautionary management of international reserves, and macroprudential policies. These adjustments helped cushion most EMs during the turbulent Global Financial Crisis of 2008 (GFC). In response to GFC, the US Federal Reserve drastically cut short-term policy interest rates and pursued unconventional monetary expansion, including quantitative easing (QE). QE policies included the Fed's purchase of longer-term bonds, aimed at flattening the yield curve.

These policies turned out to be a game-changer for the US, eurozone (EZ), and emerging markets. The sharp decline of interest rates drastically reduced the sovereign spreads of GIIPS. The negative policy interest rates of the EZ (including on most public debts of core EU countries, i.e., Belgium, France, Germany, and the Netherlands), and the sharp drop in the yields on US bonds have encouraged a global search for returns, reducing thereby the sovereign spreads of most EMs to single digits. OECD institutional investors embarked on purchasing the local currency bonds of many EMs. These developments mitigated the 'Original Sin,' allowing an increasing number of EMs to borrow both in foreign (hard) and domestic (local) currencies.[1] The resulting expansion of EMs' external debt led to an unprecedented increase in their debt to GDP ratios, bringing to the fore concerns about growing debt overhang and fragility, as well as the fiscal dominance.[2]

The patterns of sovereign spreads and the interest rate costs of local-currency and hard-currency external borrowing for EM public and private sectors have shifted in recent decades. There are possible cross effects associated with improved EM access to hard currency and local currency external borrowing, creating conditions under which the cross effects are positive or negative, possibly exacerbating fragility over time. Specifically, while shifting from hard-currency borrowing to local-currency borrowing may reduce sovereign spreads on the former, it may increase the interest rate on the latter. These effects may be non-linear, and their direction may reverse overtime for a sufficiently large debt overhang. Thus, the growing access of EMs to more elastic external borrowing in hard- and local currency imposes new debt management challenges, possibly increasing their fragility down the road and thus putting a premium on proper management of their financial and macroeconomic policies. The unconventional monetary policies adopted by the US Fed and ECB may be the 21$^{st}$-century incarnations of financial repression, as succinctly pointed out in Reinhart (2012). While financial repression may postpone adjustment of the global leverage buildup, future instability associated with the exit from debt overhang is a tail risk, heightening EM financial fragility and perhaps even triggering future EM crises. See also Diaz-Alejandro (1985) for the seminal paper on financial repression, economic vulnerability, and crisis trade-offs, and the Appendix for recent evidence of financial fragility of large EMs, focusing on BRIC (Brazil, Russia, India, and China).

---

[1] The share of local currency is estimated at 87.1 percent of total EM debt, amounting to $21.9 trillion, in 2017. Local currency debt outstanding has increased from 40 percent of GDP in the early 2010s to almost 60 percent of GDP recently (IMF, 2018).

[2] See Aizenman (2004) for further discussion on the ambiguous impact of greater financial integration of EM. Recent analysis of the impact of local currency borrowing on the spreads of local and hard currency debts include Miyajima et al. (2015), Du and Schreger (2016), Engel and Park (2019). Park et. al. (2019), and Amstad et al. (2020).

In this paper, we seek to understand the patterns of sovereign bond issuance by investigating the micro evidence grounded in bond-level information. Our primary focus is the choice of the currency denomination of the sovereign bonds. Traditionally, EMs were not able to raise funds in their own currency in the foreign markets. While some EM sovereign borrowers target domestic investors, others prioritize raising funds in the international market. In recent years, they are increasingly capable of doing so as the global demand for EM asset increases. The choice of currency denomination is then not only driven by issuers' preferences and domestic factors on the supply side, but also the demand of investors. However, a formal estimation of the causal effects of demand and supply factors requires more structural configuration and empirical details, which remains a challenge in the international finance literature. Nevertheless, our analysis aims to uncover the sailient associations in the data, highlighting the patterns found in the sovereign debts of large emerging-markets economies.

Based on the sovereign bond issuance data from the Thomson Reuters Eikon database, we focus on eight major EM sovereign borrowers, namely Brazil, China, India, Indonesia, Mexico, Russia, South Africa, and Turkey. The evidence suggests the following. First, the EM sovereign borrowers are more likely to issue local-currency bonds when the local currency appreciates particularly before the GFC. We conjecture that currency appreciation increases the prospective returns on local-currency denominated assets, which then increases the demand from investors. The stronger demand for local-currency denominated assets encourages sovereign borrowers to issue more local-currency denominated bonds. The result remains robust after we control for bond characteristics such as maturity and coupon types, country-specific economic fundamentals such as international reserves and current account balance, and global factors such as global liquidity and risk appetite.

Second, we find that inflation targeting countries, which generally have more credible monetary policies and are less likely to inflate away their public debt burden (Engel and Park, 2019), tend to issue local-currency denominated bonds before but not after GFC. This finding echoes with Hale, Jones and Spiegel (2020), who show that firms from countries with a history of stable inflation are more capable of issuing local-currency bonds, and generalize their conclusion to the context of sovereign bond issuance. The insignificant roles of inflation targeting after GFC is largely driven by the fading concerns on inflation worldwide.

Third, we find that EMs which offer higher sovereign yields after the GFC are more likely to issue local-currency denominated bonds. This finding is consistent with the global search for returns after the Fed and ECB cut the interest rate to almost zero, resulting in negative yields in some advanced markets. Finally, we also find that EM local currency bonds tend to be smaller in size, shorter in maturity, or lower in coupon rate than foreign currency bonds.

## 2. Data and Methodology

### 2.1 Data

We collect sovereign bond issuance data for eight EMs from the Thomson Reuters Eikon database for the period of 1970-2018. Our analysis focuses on Brazil, China, India, Indonesia, Mexico, Russia, South Africa and Turkey due to their relative importance in the EM bond markets and data availability. The sovereign debt of these eight EMs accounts for 89% of the total outstanding sovereign debts in EMs at the end of 2018 according to Arslanalp and Poghosyan (2014).

Our study focuses on the issuance of sovereign bond from the government's perspective, which is different from those that focuses on the domestic and foreign holdings of sovereign bonds from the investor's perspective (see for example Lane and Shambaugh, 2010, and Benetrix, Lane, and Shambaugh 2015). While knowing whether the local-currency and foreign-currency bonds are held by foreign or domestic investors are extremely valuable, we have no access to such data. To optimize debt structure, governments would consider the expected demand of sovereign bonds when choosing their currency denomination. Note that high returns and low risk attract investors, we can then form expectation on the demand of sovereign bonds issued by EM governments through bond yields, currency valuation, coupon rate and inflation risk.

### 2.2 Summary Statistics

We first present the summary statistics for local- and foreign-currency denominated bonds in Table 1. The average bond size - i.e. issue amount in USD per issuance - of foreign-currency denominated bond is larger than that of local-currency denominated bond, with the exception of China and India. Note that there are no records of foreign-currency denominated bond issuance in India in our dataset. The frequency of local-currency denominated bond issuance is higher than that of foreign-currency denominated bond issuance. If we exclude India from the sample, the ratio of local-currency bonds to the total size of bonds issued during the sample period is the highest for China (99.67%) and lowest for Russia (85%). The dominance of local-currency bonds emerges and persists in recent years. Figure 1 plots the time series of local- and foreign-currency denominated bonds issued in a particular year, with the former outpacing the latter especially after the global financial crisis across all EMs in our sample.[3]

The percentage of local-currency bond relative to all bonds issued in a particular year is larger than that based on external debt because it includes the local-currency bonds held by domestic investors. Our dataset covers bond-level currency denomination but cannot differentiate foreign investors from domestic investors. Although we do not have data on who is holding the bond, we do know the market of issuance for each bond. Traditionally bonds issued onshore mainly target domestic investors, while those issued offshore target foreign investors. However, the recent development in the debt market has enabled foreign investors to purchase more bonds issued locally and domestic investors to participate more in the offshore market. Delving into the market of issuance, we show in Panel B and C that most bonds issued onshore are denominated in local currency while most bonds issued offshore are still denominated in foreign currency. China is the only EM that had issued a larger number of local-currency bonds than foreign-currency bonds on the offshore markets in our sample. The size of local-currency bonds issued in offshore markets is not sufficient to justify the rapid growth in foreign holdings of local-currency debts documented in Arslanalp and Poghosyan (2014) and Zheng (2020). It implies rising foreign holdings of local-currency bonds issued onshore, which is also documented by Miyajima et al. (2015). Thus both onshore and offshore local-currency bond issuance contribute to the dissipation of original sin, characterized by a falling share of foreign-currency bonds in external debt. It justifies our focus on the currency denomination rather than the issuance market of sovereign bonds.

---

[3] We document similar patterns in Appendix Figure 1-3 that local-currency bond dominates both domestic and external debt when using bond-holding data from Arslanalp and Poghosyan (2014).

## 2.3 Methodology

To understand what matters for the choice of currency denomination upon sovereign bond issuance, we estimate the following baseline probit model:

$$P(D_{i,j,t} = 1) = \beta\ X_{j,t} + \gamma S_{i,j,t} + C_j + T_t + \varepsilon_{i,j,t}, \qquad (1)$$

where $D_{i,j,t} = 1$ if bond $i$ in country $j$ at period $t$ is issued in local currency and 0 otherwise. The set of key country-specific variable is $X_{j,t}$, which captures variations in the attractiveness of local-currency denominated bonds in country $j$ at period $t$. It takes the value of (i) $FX_{j,t}$, the rate of currency appreciation relative to USD in country $j$ at period $t$; (ii) $Yield_{j,t}$ is yield difference between 10-year local-currency sovereign bond in country $j$ and the US at period $t$; and (iii) $IT_{j,t}$, a dummy variable that equals 1 if country $j$ is pursuing inflation targeting at period $t$. From the investors' perspective, higher value of $Yield_{j,t}$ indicates higher return while $IT_{j,t} = 1$ means lower inflation risk for holding local-currency sovereign bonds in EMs. Given that risk-averse investors are return chasing, we expect the coefficients of $Yield_{j,t}$ and $IT_{j,t}$ to be both positive and statistically significant. The rise in currency valuation $FX_{j,t}$ increases the returns for foreign investors holding local-currency bonds, but reduces the returns for domestic investors holding foreign-currency bonds, both of which increase the demand of local-currency bonds and therefore encourage governments to issued more local-currency bonds. We therefore expect local currency appreciation to increase the likelihood of local-currency bond issuance such that the coefficient of $FX_{j,t}$ is positive and statistically significant.

We also control for a set of bond-specific variable $S_{i,j,t}$. It includes $\log(Size_{i,j,t})$, the logarithm of the issued amount of bond $i$ in country $j$ at period $t$, $\log(Maturity_{i,j,t})$, the logarithm of the maturity of bond $i$ in country $j$ at period $t$, and $Zero_{i,j,t}$, and a dummy that equals to one if the bond $i$ in country $j$ at period $t$ is a zero-coupon bond. These bond characteristics are determined by the issuers and capture the supply-side information in the bond market. The issuance of each sovereign bond is driven by a specific funding need. The bond size, maturity and coupon rate reflect such funding needs of governments. Governments need to consider the market demand of their bonds upon issuance. The coupon rate captures by $Zero_{i,j,t}$ reflects the micro-level repayment arrangement, which could also be an important determinant of bond yields that affect the demand of the bonds. The variable $C_j$ and $Y_t$ are country and year fixed effects, respectively, and $\varepsilon_{i,j,t}$ is the error term clustered by country.

## 3. Empirical Results

We explore the likelihood of issuing local-currency denominated bond using bond-level data in this section. We hypothesize that EMs are increasingly capable of issuing local-currency denominated bond because of (i) rising currency valuations that deliver additional returns to local-currency bonds, which encourages investors to hold local-currency bonds; (ii) risk-on exposures to EM as investors seek high yields since sovereigns bond in advanced markets are offering very low or negative yields; and (iii) inflation targeting that increases the credibility of EM's monetary policy and reduces the probability of currency debasement (Engel and Park, 2019).

**3.1 Baseline Results**

Table 2 summarizes the baseline probit regression results. Column 1 suggests that the appreciation of the local currency, characterized by a positive $FX_{j,t}$, is significantly associated with a higher probability of issuing a sovereign bond in local currency. Based on the marginal effects reported in Appendix Table 1, a 10% appreciation in local currency relative to USD is associated with 1.8% higher probability of issuing local-currency bonds. Appendix Table 2 shows that the positive association between currency valuation and the likelihood of local-currency bond issuance remains the same when we replace $FX_{j,t}$ with its one-year lag. Due to the autocorrelation in currency valuation, controlling for additional lags does not add value to the model performance. The finding that local-currency bond issuances are more likely when the local currency appreciates provides empirical support to the theoretical results in Ottonello and Perez (2019). Given that lower credit spreads in local-currency bonds encourage governments to issue more local-currency bonds, our finding is also consistent with Hofmann, Shim and Shin (2019), who document that local currency appreciation reduces credit risk premium and subsequently sovereign bond spreads in EMs.

Higher yield in local-currency sovereign bonds increases their attractiveness. The rising demand by investors is expected to encourage more issuance of local-currency bonds. However we find no evidence to support such a hypothesis based on the full sample. The result in column 2 shows that, the yield spread between country $j$ and the US, $Yield_{j,t}$, does not seem to change the likelihood of local-currency denominated bond issuance. Investors' preference over the risk-return tradeoff may vary over time, especially when the market shifts to the new normal of ultra low interest rate. We take into account of the changing market environment in the next section to further explore the role of yield searching on the currency denomination of sovereign bond.

Inflation targeting reduces the inflation risk in EMs, which is expected to increase the demand of local-currency bonds that enables governments to issue more local-currency bonds. Column 3 shows that the coefficient of inflation targeting dummy $IT_{j,t}$ is positive but not statistically significant. It appears that inflation targeting increases the probability of issuing local-currency denominated bonds. The result is, however, not statistically significant. The roles of inflation risk may have shifted after a decade's stable inflation in most countries including EMs. We account for the fading concerns on inflation after GFC in the next section.

The result in column 4 shows that controlling for inflation targeting does not affect the roles of currency valuation on the probability of issuing local-currency bonds.[4] Consistent with the summary statistics in Table 1, bonds that are smaller in size, shorter in maturity, or lower in coupon rate are more likely to issue in local currency. Similar evience on the negative association between bond size and the likelihood of local-currency bond issuance is found in the context of corporate bonds by Hale, Jones, and Spiegel (2020).

## 3.2 Robustness Checks

We check whether the positive relation between currency appreciation and the likelihood of issuing local-currency bond is robust in this section. To address the concern of omitted variables, we further control for a number of domestic variables that could possibly affect the choice of bond denomination. A strong current account (CA) balance may increase a country's capacity to repay the debt and reduce the default risk. The result in Column 1 of Table 3 suggests that higher CA balance may enable the government to issue more local-currency bond. However, the positive relation between currency appreciation and the likelihood of issuing local-currency bond is not driven by CA balance. The coefficient of FX remains positive and statistically significant after controlling for CA balance (see Column 1).

---

[4] We experiment with CIP deviation but the results are not significant, possibly due to the lack of hedging tools to hedge typically long-term sovereign bonds.

Column 2 of Table 3 accounts for the effect of domestic investment, showing its positive and significant association with local-currency bond issuance. Note that incomes generated by domestic investments are in local currency, which tends to depreciate in economic recessions and appreciate in booms. Financing domestic with local-currency bond provides a natural hedge to income risk and mitigates the problem of current mismatch (Ottonello and Perez, 2019, Eichengreen, Hausmann and Panizza, 2007). Thus governments are motivated to issue more local-currency bond issuance to finance stronger domestic investment.

Column 3 of Table 3 shows that countries with higher GDP per capita growth is more likely to issue foreign-currency bonds. GDP per capita growth is typically faster for less developed economies, whose financial market is less matured and more volatile. The high risk associated with the local-currency denominated assets may disable these EMs from raising funds in their own currency, and motivates governments to issue foreign-currency bonds to meet their funding needs. Column 4 of Table 3 shows that countries with a higher international reserve are more likely to issue foreign-currency bonds. Reserve provides an efficient buffer against foreign exchange rate shock, which mitigates the risk of foreign-currency debt and encouarges the issuance of foreign-currency bonds.

The results remain robust when we control for all the domestic factors mentioned above (see Column 5 of Table 3). Despite the fact that the relation between these domestic factors, which affect the choice of bond currency denomination, and currency valuation may affect our results, the positive relation between currency appreciation and the likelihood of local-currency bond issuance remain robust after controlling for these domestic factors. Appendix Table 3 show that even after controlling for the fiscal position and government expenditure, the key result remains robust. Note however that the coefficients of fiscal position and government expenditure are not statistically significant.

In the baseline regression, we control for year fixed effects which absorb any global factors that affect the emerging markets' choice of bond currency denomination. It would be interesting to see how the global financial cycle and international financial market conditions affect EM's bond issuance behavior. Table 5 reports the estimation results that replace year fixed effects with various global factors to check the robustness of our main findings and understand their role in the choice of sovereign bond currency denomination. Regardless of the specific global factor that we control for, the positive relation between currency appreciation and the likelihood of local-currency bond issuance remains robust. We find that the global risk appetite indicator VIX, measured by the log return of the CBOE volatility index, has little influence on the choice of bond currency denomination. If global liquidity is abundant, as indicated by a lower value of Ted, the interest difference between 3-M LIBOR based on US dollars and 3-M US Treasury bill, EMs are more likely to issue local-currency bonds. When the global liquidity is constrained and the market is in a risk-off mode, investors prefer USD assets over EM currency asses, which reduces the demand of EM local-currency bonds and subsequently discourages governments from issuing local-currency bonds.

By the same logic, the shock of higher oil prices, measured by the log return in the crude oil price, would push global financial markets to a risk-off mode, which would trigger more demand for USD-denominated assets and less demand for local-currency EM bonds. There is no evidence that global policy uncertainty, measured by the log return of the global Economic Policy Uncertainty Index, affects the EMs' choice of bond currency denomination. Global factors add new information on the bond currency denomination, but they do not affect our main findings.

One may argue that the market environment accommodative for issuing local currency bonds tends to be more favorable for issuing short maturity and smaller bonds as well.[5] To address such a concern, we control for factors that may simultaneously affect the choice of currency denomination, size and maturity of bonds. Throughout all regressions in Table 3 and 4, the coefficients of log(Size) and log(Maturity) are negative and statistically significant, while that of *Zero* is negative and statistically significant, which is consistent with the basline result that smaller, shorter, and lower-coupon bonds are more likely to be issued in local currency. It therefore mitigates the concern that our baseline results on the relation between bond characteristics and the likelihood of local-currency bond issuance is not driven by the common factors.

Running probit regressions with fixed effects can lead to biased estimation because of the incidental parameters problem. We follow Hahn and Newey (2004) to use an analytical bias correction motivated by large time periods and re-estimate the probit model. The estimation results presented in Appendix Table 4 shows that the positive association between currency appreciation and the likelihood of local-currency bond issuance remain significant after correcting the bias caused by incidental parameters problem.

---

[5] We thank an anonymous reviewer for pointing out this alternative explanation.

**3.3 Heterogeneity analysis**

Du and Tepper (2016) and Du, Tepper and Verdelhan (2018) show that the international debt market shows different patterns after the global financial crisis (GFC). In particular, covered interest rate parity (CIP) no longer holds after GFC in both AMs and EMs. The deviation from CIP may change bond issuers' preferences for local vis-a-vis foreign currency denomination. To explore whether sovereign bond issuance patterns changed after GFC, we extend Equation (1) to account for the interaction between the key independent variables and GFC, a dummy that equals 1 after 2007 and 0 otherwise.

Table 5 shows that the GFC indeed reshaped bond issuance patterns.[6] The result in column 1 suggests that before GFC a government is more likely to issue local-currency denominated bonds when the local currency appreciates, the relation is reversed after GFC. The coefficient of the interaction between FX and GFC is negative and statistically significant and its magnitude exceeds that of the coefficient of FX. The sum of the coefficients of FX and its interaction with GFC is negative (-23) and statistically significant at the level of 1% ($\chi^2 = 10.538$). It suggests that, after GFC, domestic currency appreciation is associated with a lower probability of issuing local-currency bonds. The result could potentially be driven by the concern on the potential risk underlying the currency appreciation caused by quantitative easing in the US and the potential reversal associated with i.e. sudden stops, which downplays the attractiveness of local-currency denominated bond.

---

[6] We find similar evidence in Sppendix Table 5 when applying the linear probability model.

The association between $Yield_{j,t}$ and the likelihood of local-currency bond issuance varies before and after GFC, as shown in Column (2) of Table 5. It turns out that the previous result of the limited association between yield difference between local and US bond and the likelihood of local-currency bond issuance is driven by mixed effects in different market regimes. Before GFC, the government is less likely to issue local-currency bond when $Yield_{j,t}$ is positive - i.e. it is more costly to issue local-currency bonds. The results reflect rationally minimizing funding costs by raising funds in a less costly way. Summing the coefficients of $Yield_{j,t}$ and its interaction with GFC yields a positive coefficient (0.88) that is statistically significant at 1%. It suggests that, after GFC, the government is more likely to issue local-currency bonds when local yields are higher than the US, which is consistent with the deviation from CIP in emerging markets documented by Du and Tepper (2016).

It is puzzling that EMs issue more local-currency denominated bonds even though they are more expensive. A possibility is the growing desire of EMs to strengthen their resilience to external shocks through risk-sharing and currency matching. The benefits from such strengthening may well exceed the additional funding costs that exceed the USD-denominated bonds. EMs were traditionally unable to raise funds in their own currency, which exposed them to substantial external shocks, especially when the USD appreciates relative to local currency. As investors search for yield in the post-GFC low interest environment and EM inflation fell in recent years, the demand for EM-currency bonds rose, which allowed EMs to issue local-currency bond more easily. It seems that the demand of international investors for local-currency denominated EM bonds dominates the decision-making process on the choice of issuance currency after GFC. Our results suggest that the low interest-environment after GFC provides an opportunity for EMs to issue bonds denominated in their own currency, especially bonds that offer higher yield relative to US bonds.

We find that inflation targeting increases the likelihood of issuing local-currency bonds before but not after GFC (see Column 3 of Table 5). Again, such mixed results are driving the insignificant relation between inflation targeting and the likelihood of local-currency bond issuance in Table 2. A country can increase its money supply significantly to inflate away the debt burden. EM inflation was thus a serious concern for international investors. Adopting inflation targeting restrains central banks from printing money to erode their debt, and thereby increases the credibility of their monetary policy. As such, inflation targeting mitigates the perceived risk of investing in EM's local-currency denominated bond and attracts more demand from investors. Inflation targeting therefore enables EMs to issue local-currency bonds more easily. Our results provide similar evidence with Hale, Jones and Spiegel (2020), who document that a history of stable inflation enables firms to issue more local-currency bonds, in the context of sovereign bonds. Our estimation results before GFC confirm that inflation targeting enables EMs to issue more local-currency bonds, which is also consistent with the theoretical implication in Engel and Park (2019).

However, we find that inflation-targeting countries no longer enjoy such a privilege after GFC, when inflation rate remained low despite massive quantitative easing in major advanced economies. When low inflation becomes the norm, commitment to keep inflation low is no longer as valuable. There are two possible explanations for the negative relation between inflation targeting and the likelihood of issuing local-currency bond. First, inflation-targeting countries offer a lower yield compared to non-inflation targeting countries. Second, currency appreciates more in inflation-targeting countries than non-inflation-targeting countries after GFC. The result in column

(4) shows that, after controlling for currency valuation and yield difference, the role of inflation targeting fades away. It suggests that either currency valuation or yield difference or both have absorbed the effects of inflation targeting.

## 4. Concluding Remarks

Both advanced countries and emerging markets have substantially expanded their public-sector borrowing as a share of GDP since the global financial crisis. This trend was driven by the secular decline of risk-free interest-rates, a process that was magnified by the unconventional monetary expansions of the US Fed and ECB. The GFC led to public sector bailouts of financial institutions, and the large-scale socialization of their private losses. Quantitative expansion (QE) and other expansionary monetary policies resulted in the secular decline of interest rates, and growing fiscal dominance. This may be the modern incarnation of financial repression, as articulated by Reinhart (2012).[7] According to this view, the post-GFC monetary policies of the US and eurozone drastically reduced the cost of servicing sovereign debt, in ways that reflect political economy factors. These effects propagated the unprecedented post-GFC EM leverage buildup, funded this time by both hard and domestic currency external borrowing.

---

[7] "One of the main goals of financial repression is to keep nominal interest rates lower than would otherwise prevail. This effect, other things being equal, reduces governments' interest expenses for a given stock of debt and contributes to deficit reduction. However, when financial repression produces negative real interest rates and reduces or liquidates existing debts, it is a transfer from creditors (savers) to borrowers and, in some cases, governments. This amounts to a tax that has interesting political-economy properties. Unlike income, consumption, or sales taxes, the repression tax rate is determined by factors such as financial regulations and inflation performance, which are opaque—if not invisible—to the highly politicized realm of fiscal policy. Given that deficit reduction usually involves highly unpopular spending cuts and/or tax increases, the stealthier financial-repression tax may be a more politically palatable alternative. Key factors underlying the high incidence of negative real interest rates after the crisis are aggressively expansive stance of monetary policy and heavy central bank intervention in many advanced and emerging economies. This raises the broad question of whether current interest rates are more likely to reflect market conditions or whether they are determined by the actions of large official players in financial markets. A large role for nonmarket forces in interest-rate determination is a central feature of financial repression." Reinhart (2012).

The history of financial repression suggests that it may act as a pain killer, delaying the adjustment required to address the debt build-up, and providing the illusion of stability. With luck, a gradual exit strategy from the debt overhang will allow the US and eurozone to spread the adjustment over a decade or more, possibly by means of higher economic growth and inflation. Indeed, this was the post-WWII exit strategy of the US during 1945-1955, eventually reducing the public debt to GDP from about 110% to about 50% [Aizenman and Marion (2011)]. Yet exits from the debt overhang may also induce deflationary spells and lower growth rates [Lo and Rogoff (2012), Reinhart et al. (2012)]. Even a relatively fast exit from higher inflationary and leverage spells by OECD countries may destabilize EMs with less developed financial markets and relatively small tax bases.[8] The exposure of EM to this tail risk associated with the advanced economies' exit from debt overhangs remains a source of potential EM fragility and may even trigger future crises, possibly well before the actual OECD countries exit from the present debt overhang.

Our empirical analysis of sovereign bond issuance data from eight major emerging markets (EMs) in 1970-2018 lends further weight to such concerns. Our analysis is centered on delving into micro data to identify the key determinants of local-currency sovereign bond issuance. We find that EM bonds which are smaller in size, shorter in maturity, or lower in coupon rate are more likely to be issued in local currency. Our evidence indicates that there have been structural changes in the determinants of EMs' local-currency sovereign bond issuance since GFC. More specifically, we find that EMs are more likely to issue local-currency sovereign bonds if (1) domestic currencies appreciate, but only before GFC, (2) the monetary policy regime is inflation targeting, before but not after GFC, and (3) bonds offer a higher yield, but only after GFC. Taken together, these findings suggest that even EMs with less robust fundamentals are more able to issue local-currency sovereign bonds in the post-GFC period.

---

[8] To recall, Paul Volcker's disinflationary policies in the US during 1979-1983 triggered the EM lost decade. See Aizenman et al. (2019) for further analysis of EM fragility and fiscal space.

In addition, the devastating COVID-19 health and economic crisis is likely to significantly increase the borrowing requirements of EM public sectors. Therefore, the risk of financial turbulence in EMs is likely to remain substantial despite their overcoming the Original Sin. EMs' enhanced ability to borrow abroad in local currency integrates them more closely into the global financial system. Closer financial integration yields significant benefits for EMs but also renders their financial stability more vulnerable to events in advanced economies, such as their exit from their debt overhang.. Finally, future data will allow us to test and identify structural changes associated with the COVID-19 pandemic and its aftermath.

Figure 1. Local- and foreign-currency bond issuance comparison in annual total value

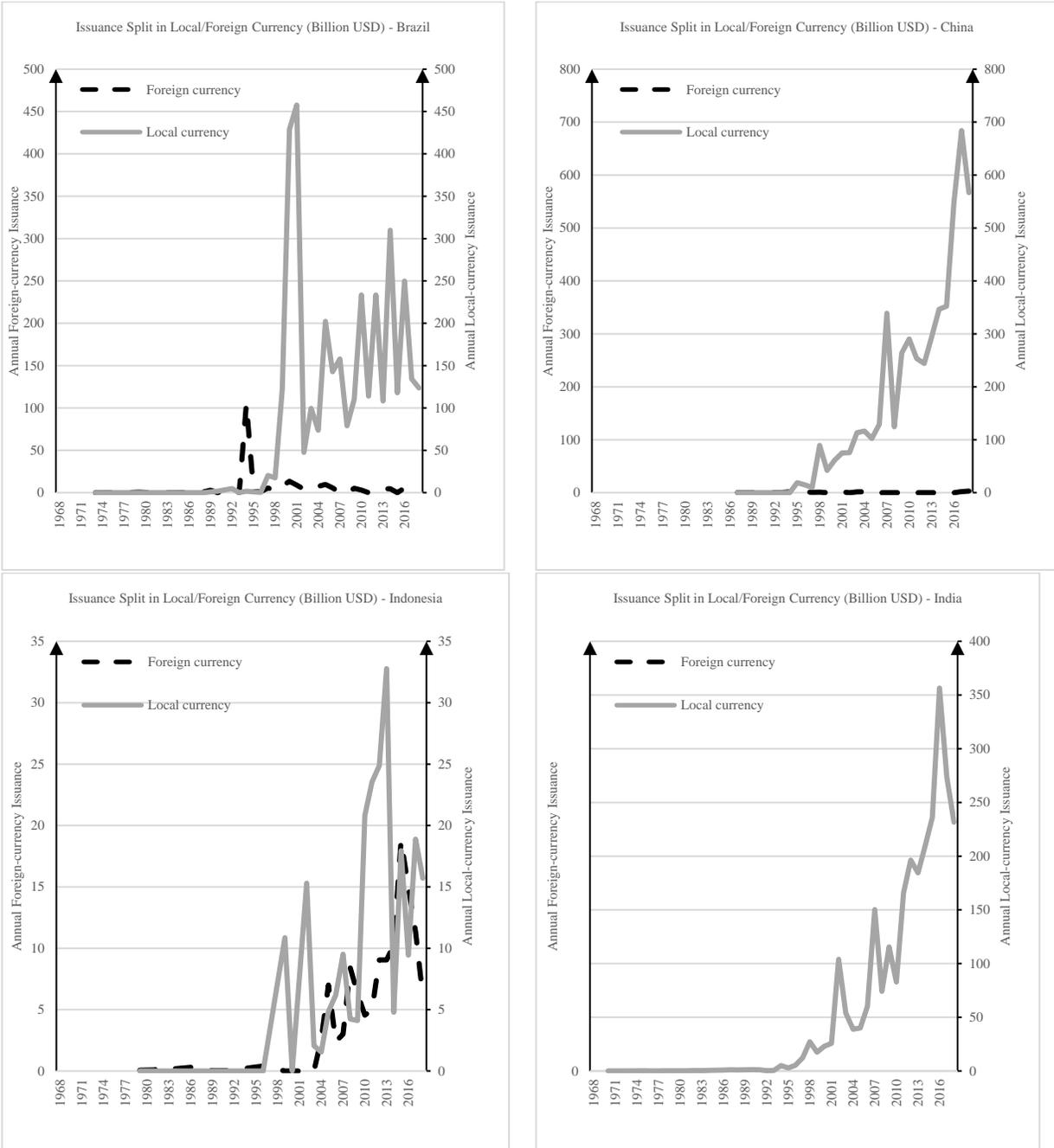

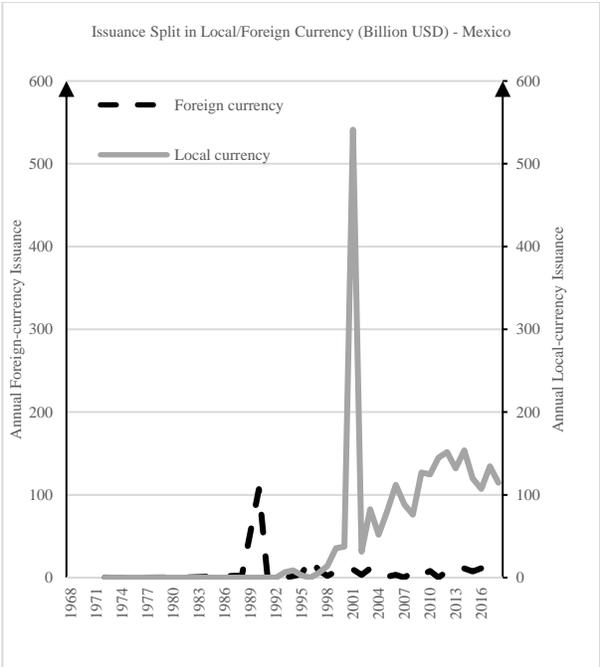
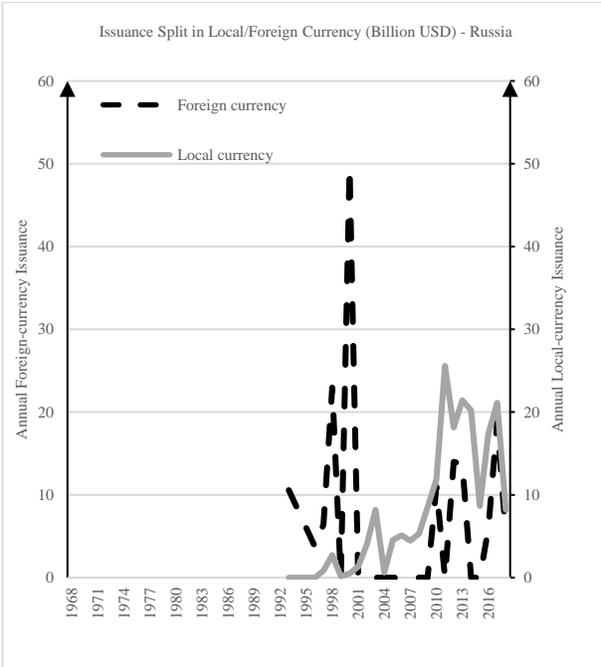
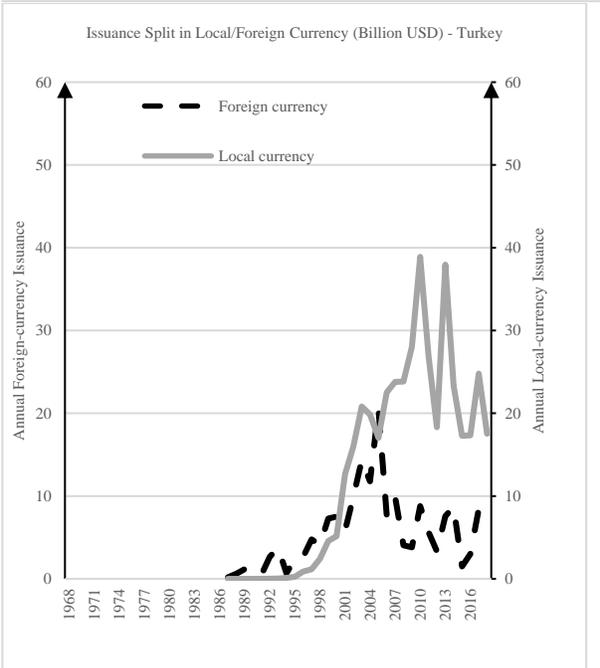
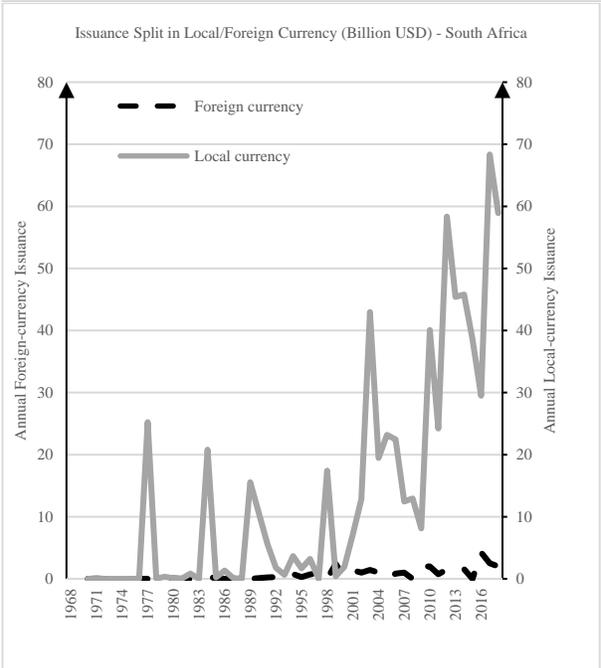

**Table 1. Summary statistics**

This table reports the average bond size in million USD and the number of bond issuances, denominated in local and foreign currency respectively, for eight emerging markets during 1970-2018. Local/Total is the total amount (number) of bond issued in local currency relative to the total amount (number) of bond issuances in the sample.

| Country | Country ISO | Average bond size (Million $) | | Number of Bond | |
|---|---|---|---|---|---|
| | | Local currency | Foreign currency | Local currency | Foreign currency |
| **Panel A: All bonds** | | | | | |
| Brazil | BRA | 570.08 | 1981.94 | 6302 | 110 |
| China | CHN | 4270.22 | 548.24 | 1129 | 29 |
| Indonesia | IDN | 621.67 | 1579.02 | 365 | 76 |
| India | IND | 1132.28 | NA | 2388 | 0 |
| Mexico | MEX | 1408.93 | 1534.74 | 1764 | 180 |
| Russia | RUS | 881.67 | 4380.77 | 215 | 37 |
| Turkey | TUR | 529.35 | 941.66 | 796 | 193 |
| South Africa | ZAF | 392.74 | 694.84 | 1645 | 46 |
| | Total | 1033.06 | 1499.2 | 14604 | 671 |
| **Panel B: Bonds issued onshore** | | | | | |
| Brazil | BRA | 570.36 | 0 | 6298 | 0 |
| China | CHN | 4828.77 | 0 | 1062 | 0 |
| Indonesia | IDN | 622.98 | 300.47 | 365 | 5 |
| India | IND | 1132.28 | 0 | 2388 | 0 |
| Mexico | MEX | 1409.72 | 147.24 | 1763 | 13 |
| Russia | RUS | 917.73 | 2297 | 214 | 2 |
| Turkey | TUR | 529.35 | 878.56 | 796 | 56 |
| South Africa | ZAF | 408.79 | 1.96 | 1644 | 1 |
| | Total | 1061.71 | 743.01 | 14530 | 77 |
| **Panel C: Bonds issued offshore** | | | | | |
| Brazil | BRA | 930.16 | 1981.94 | 4 | 110 |
| China | CHN | 414.62 | 566.03 | 67 | 28 |
| Indonesia | IDN | 0 | 1670.67 | 0 | 71 |
| Mexico | MEX | 1.68 | 1642.75 | 1 | 167 |
| Russia | RUS | 2592.3 | 4499.84 | 1 | 35 |
| Turkey | TUR | 0 | 967.46 | 0 | 137 |
| South Africa | ZAF | 0 | 710.24 | 0 | 45 |
| | Total | 467.05 | 1600.03 | 73 | 593 |

**Table 2: Baseline probit regression results**
This table reports the estimation results from the following probit regression:

$$P(D_{i,j,t} = 1) = \beta \ X_{j,t} + \gamma S_{i,j,t} + C_j + T_t + \varepsilon_{i,j,t} \ (1),$$

where $D_{i,j,t} = 1$ bond $i$ in country $j$ at period $t$ is issued in local-currency. The key country-specific variable is $X_{j,t}$, which takes the value of (i) $FX_{j,t}$, the currency appreciation of country $j$ at period $t$ relative to USD; (ii) $Yield_{j,t}$ 10-year sovereign bond yield difference between country $j$ and US at period $t$; and (iii) $IT_{j,t}$, a dummy variable that equals 1 if country $j$ is pursuing inflation targeting at period $t$. The vector of bond-level control variable $S_{i,j,t}$ covers (i) log ($Size_{i,j,t}$), the logarithm of the issued amount of bond $i$ in country $j$ at period $t$; (ii) log ($Maturity_{i,j,t}$), the logarithm of the maturity of bond $i$ in country $j$ at period $t$; and (iii) $Zero_{i,j,t}$, a dummy that equals to one if the bond $i$ in country $j$ at period $t$ is a zero-coupon bond. The variable $C_j$ and $Y_t$ are country and year fixed effects respectively. Standard errors reported in the parenthesis are clustered by country. ***, ** and * denote significance level at 1%, 5% and 10%.

|  | \multicolumn{4}{c}{Dependent variable: $P(D_{i,j,t} = 1)$} | | | |
| --- | --- | --- | --- | --- |
|  | (1) | (2) | (3) | (4) |
| FX | 4.236*** |  |  | 4.287*** |
|  | (1.464) |  |  | (1.462) |
| Yield |  | -0.018 |  |  |
|  |  | (0.027) |  |  |
| IT |  |  | 0.077 | 0.113 |
|  |  |  | (0.110) | (0.111) |
| log(Size) | -0.253*** | -0.182*** | -0.245*** | -0.253*** |
|  | (0.016) | (0.020) | (0.015) | (0.016) |
| log(Maturity) | -0.384*** | -0.324*** | -0.380*** | -0.386*** |
|  | (0.038) | (0.052) | (0.038) | (0.038) |
| Zero | 1.055*** | 1.909*** | 1.101*** | 1.061*** |
|  | (0.112) | (0.296) | (0.112) | (0.111) |
| Constant | 8.574*** | 7.703*** | 13.672 | 8.567*** |
|  | (0.518) | (1.045) | (10,696.730) | (0.518) |
| Observations | 15,072 | 12,297 | 15,271 | 15,072 |
| Log Likelihood | -1,191.020 | -637.551 | -1,274.029 | -1,190.495 |
| Country fixed effects? | Yes | Yes | Yes | Yes |
| Year fixed effects? | Yes | Yes | Yes | Yes |

**Table 3: Controlling for additional domestic factors**

This table reports the estimation results from the probit regression:

$$P(D_{i,j,t} = 1) = \beta\ FX_{j,t} + \gamma S_{i,j,t} + \tau DF_{j,t} + C_j + T_t + \varepsilon_{i,j,t},$$

where $D_{i,j,t} = 1$ bond $i$ in country $j$ at period $t$ is issued in local-currency. $FX_{j,t}$ is the currency appreciation of country $j$ at period $t$ relative to USD. The domestic factor $DF_t$ includes (i) CA Balance, the current account balance normalized by GDP; (ii) Investment, the domestic investment normalized by GDP; (iii) Growth, the GDP per capita growth rate; and (iv) Reserve, the international reserve normalized by GDP. The vector of bond-level control variable $S_{i,j,t}$ covers (i) log ($Size_{i,j,t}$), the logarithm of the issued amount of bond $i$ in country $j$ at period $t$; (ii) log ($Maturity_{i,j,t}$), the logarithm of the maturity of bond $i$ in country $j$ at period $t$; and (iii) $Zero_{i,j,t}$, a dummy that equals to one if the bond $i$ in country $j$ at period $t$ is a zero-coupon bond. The variable $C_j$ and $Y_t$ are country and year fixed effects respectively. Standard errors reported in the parenthesis are clustered by country. \*\*\*, \*\* and \* denote significance level at 1%, 5% and 10%.

|  | \multicolumn{5}{c}{Dependent variable: $P(D_{i,j,t}=1)$} | | | | |
|---|---|---|---|---|---|
|  | (1) | (2) | (3) | (4) | (5) |
| FX | 5.048\*\* | 4.166\* | 5.125\*\*\* | 4.687\*\* | 9.507\*\*\* |
|  | (2.317) | (2.289) | (1.516) | (2.370) | (1.813) |
| log(Size) | -0.214\*\*\* | -0.219\*\*\* | -0.261\*\*\* | -0.237\*\*\* | -0.194\*\*\* |
|  | (0.018) | (0.018) | (0.017) | (0.019) | (0.016) |
| log(Maturity) | -0.348\*\*\* | -0.375\*\*\* | -0.365\*\*\* | -0.379\*\*\* | -0.206\*\*\* |
|  | (0.039) | (0.041) | (0.039) | (0.041) | (0.036) |
| Zero | 0.976\*\*\* | 1.008\*\*\* | 1.091\*\*\* | 0.991\*\*\* | 1.298\*\*\* |
|  | (0.115) | (0.117) | (0.115) | (0.116) | (0.117) |
| CA Balance | 0.036\*\* |  |  |  | 0.025\*\* |
|  | (0.016) |  |  |  | (0.011) |
| Investment |  | 0.007\*\*\* |  |  | 0.018\*\*\* |
|  |  | (0.002) |  |  | (0.001) |
| Growth |  |  | -0.026\*\* |  | -0.054\*\*\* |
|  |  |  | (0.012) |  | (0.010) |
| Reserve |  |  |  | -0.542\* | -0.369\*\* |
|  |  |  |  | (0.322) | (0.165) |
| Constant | 7.868\*\*\* | 7.872\*\*\* | 8.590\*\*\* | 8.933\*\*\* | 5.754\*\*\* |
|  | (0.830) | (0.981) | (0.530) | (0.984) | (0.406) |
| Observations | 14,517 | 13,319 | 14,563 | 13,918 | 12,403 |
| Log Likelihood | -1,093.170 | -1,026.425 | -1,129.459 | -993.694 | -1,087.018 |
| Country fixed effects? | Yes | Yes | Yes | Yes | Yes |
| Year fixed effects? | Yes | Yes | Yes | Yes | Yes |

**Table 4: Controlling for global factors**

This table reports the estimation results from the following probit regression

$$P(D_{i,j,t} = 1) = \beta\ FX_{j,t} + \gamma S_{i,j,t} + C_j + GF_t + \varepsilon_{i,j,t},\quad (1)$$

where $D_{i,j,t} = 1$ bond $i$ in country $j$ at period $t$ is issued in local-currency. $FX_{j,t}$ is the currency appreciation of country $j$ at period $t$ relative to USD. The global factor $GF_t$ includes (i) VIX, the log return of the CBOE volatility index; (ii) Ted Spread, the interest difference between 3-M LIBOR based on US dollars and 3-M US Treasury bill; (iii) Oil Price Shock, the log return in the crude oil price; and (iv) Policy Uncertainty, the log return of the global Economic Policy Uncertainty Index. The vector of bond-level control variable $S_{i,j,t}$ covers (i) log ($Size_{i,j,t}$), the logarithm of the issued amount of bond $i$ in country $j$ at period $t$; (ii) log ($Maturity_{i,j,t}$), the logarithm of the maturity of bond $i$ in country $j$ at period $t$; and (iii) $Zero_{i,j,t}$, a dummy that equals to one if the bond $i$ in country $j$ at period $t$ is a zero-coupon bond. The variable $C_j$ and $Y_t$ are country and year fixed effects respectively. Standard errors reported in the parenthesis are clustered by country. ***, ** and * denote significance level at 1%, 5% and 10%.

|  | Dependent variable: $P(D_{i,j,t} = 1)$ | | | | |
|---|---|---|---|---|---|
|  | (1) | (2) | (3) | (4) | (5) |
| FX | 9.814*** | 9.696*** | 10.343*** | 4.641** | 4.725** |
|  | (1.002) | (1.001) | (1.016) | (1.834) | (1.941) |
| log(Size) | -0.165*** | -0.165*** | -0.168*** | -0.157*** | -0.160*** |
|  | (0.011) | (0.011) | (0.012) | (0.014) | (0.014) |
| log(Maturity) | -0.294*** | -0.302*** | -0.301*** | -0.279*** | -0.296*** |
|  | (0.033) | (0.033) | (0.033) | (0.037) | (0.038) |
| Zero | 1.062*** | 1.048*** | 1.054*** | 1.201*** | 1.170*** |
|  | (0.106) | (0.106) | (0.105) | (0.130) | (0.129) |
| VIX | -0.243 |  |  |  | 1.340 |
|  | (0.833) |  |  |  | (1.325) |
| Ted Spread |  | -0.163* |  |  | -0.280** |
|  |  | (0.095) |  |  | (0.117) |
| Oil Price Shock |  |  | -2.596*** |  | -1.110 |
|  |  |  | (0.897) |  | (1.049) |
| Policy Uncertainty |  |  |  | 0.550 | -0.051 |
|  |  |  |  | (0.930) | (1.209) |
| Constant | 7.198*** | 7.338*** | 7.335*** | 6.934*** | 7.257*** |
|  | (0.355) | (0.364) | (0.361) | (0.419) | (0.439) |
| Observations | 15,072 | 15,072 | 15,072 | 14,162 | 14,162 |
| Log Likelihood | -1,353.824 | -1,352.496 | -1,349.734 | -1,080.347 | -1,076.830 |
| Country fixed effects? | Yes | Yes | Yes | Yes | Yes |
| Year fixed effects? | No | No | No | No | No |

**Table 5: Heterogeneity before and after Global Financial Crisis (GFC)**
This table reports the estimation results from the following probit regression

$$P(D_{i,j,t} = 1) = \beta\ X_{j,t} + \vartheta\ X_{j,t} * GFC_t + \gamma S_{i,j,t} + C_j + T_t + \varepsilon_{i,j,t}, \quad (1)$$

where $D_{i,j,t} = 1$ bond $i$ in country $j$ at period $t$ is issued in local-currency. $GFC_t$ is a dummy variable that equals to 1 after 2007 and 0 otherwise. The country-specific variable is $X_{j,t}$, which takes the value of (i) $FX_{j,t}$, the currency appreciation of country $j$ at period $t$ relative to USD; (ii) $Yield_{j,t}$ 10-year sovereign bond yield difference between country $j$ and US at period $t$; and (iii) $IT_{j,t}$, a dummy variable that equals 1 if country $j$ is pursuing inflation targeting at period $t$. The vector of bond-level control variable $S_{i,j,t}$ covers (i) $\log(Size_{i,j,t})$, the logarithm of the issued amount of bond $i$ in country $j$ at period $t$; (ii) $\log(Maturity_{i,j,t})$, the logarithm of the maturity of bond $i$ in country $j$ at period $t$; and (iii) $Zero_{i,j,t}$, a dummy that equals to one if the bond $i$ in country $j$ at period $t$ is a zero-coupon bond. Chi-squared and the p-value are from the test of null hypothesis that the sum of FX (yield, IT) and its interaction with GFC is 0. The variable $C_j$ and $Y_t$ are country and year fixed effects respectively. The wald test oStandard errors reported in the parenthesis are clustered by country. ***, ** and * denote significance level at 1%, 5% and 10%.

| | (1) | (2) | (3) |
|---|---|---|---|
| FX | 5.251*** | | |
| | (1.477) | | |
| Yield | | -0.093*** | |
| | | (0.034) | |
| IT | | | 0.304** |
| | | | (0.125) |
| FX*GFC | -28.594*** | | |
| | (6.511) | | |
| Yield*GFC | | 0.162*** | |
| | | (0.050) | |
| IT*GFC | | | -0.720*** |
| | | | (0.194) |
| GFC | 0.738** | -1.451 | -4.031 |
| | (0.332) | (0.907) | (10,696.720) |
| log(Size) | -0.255*** | -0.181*** | -0.242*** |
| | (0.016) | (0.020) | (0.015) |
| log(Maturity) | -0.395*** | -0.323*** | -0.380*** |
| | (0.038) | (0.052) | (0.038) |
| Zero | 1.074*** | 1.954*** | 1.103*** |
| | (0.113) | (0.300) | (0.111) |
| Constant | 8.729*** | 8.273*** | 13.579 |

|  | (0.522) | (1.051) | (10,696.720) |
| --- | --- | --- | --- |
| Observations | 15,072 | 12,297 | 15,271 |
| Log Likelihood | -1,181.284 | -631.928 | -1,266.973 |
| Country fixed effects? | Yes | Yes | Yes |
| Year fixed effects? | Yes | Yes | Yes |
|  | FC+FC*GFC | Yield+Yield*GFC | IT+IT*GFC |
| Chi-squared | 10.538 | 11.828 | 2.955 |
| p-value | 2.267×10-5 *** | 7.381×10-6 *** | 0.052** |

Online Appendix

A1. Introduction

In addition to the Fed policies, the eurozone sovereign debt crisis and the ensuing sovereign debt crises of Greece, Ireland, Italy, Portugal, and Spain (GIIPS) induced the ECB to adopt Mario Draghi's version of unconventional monetary policies. *Draghi's ECB tenure: Saving the euro, faltering on inflation*, FT 10/20/19 concisely summarized these policies. Mario Draghi, then the ECB President, "expanded the ECB's policy toolbox to include generous subsidised lending to banks to help shore up their balance sheets, negative rates to lower borrowing costs and sovereign bond purchases to bring down the market interest rates faced by the bloc's most troubled economies." Consequently, "The ECB has subsequently accumulated €2.6tn of assets, including nearly a quarter of member states' outstanding bonds. Critics in northern Europe complained that these programmes were beyond the bank's mandate, while others warned that the negative side-effects outweighed the benefits." "In the eurozone, government bond yields measure investors' perception of risk — the more likely the markets think it is that a country will crash out of the bloc, the wider the spread between its yields and those of Germany, the single currency's largest economy. During Mr Draghi's tenure, peripheral countries' spreads shot up to historic highs as investors became fearful that they would be unable to finance their rising debt levels or stimulate their struggling economies. The bloc's banks are large holders of their home nations' debt, so the sovereign debt crisis soon evolved into a banking crisis, and that in turn hit lending to households and businesses. The subsequent retrenchment in eurozone bond spreads demonstrated that Mr Draghi's use of unconventional monetary policy had worked, economists say. 'It is widely agreed that [his] pledge to make the ECB the de facto lender of last resort to governments was the key to arresting the euro crisis,' said Christian Odendahl, chief economist at the Centre for European Reform."

On the issue of fiscal dominance, this possibility can arise when debt/GDP constrains the conduct of monetary policy by forcing the central bank to pay attention to reducing the costs of servicing the public debt and external debt [see Blanchard (2004) for an overview, and also Ahmed et al. (2019) for a recent cross-country evidence in the context of inflation targeting regimes and fiscal dominance]. The distinction between fiscal and monetary dominance regimes is due to Sargent and Wallace (1981). If the government adjusts the primary deficit to limit debt accumulation, the central bank is not forced to inflate away the debt, allowing the central bank to focus on inflation targeting, in line with monetary dominance. However, long periods of large fiscal deficits and high public debt-to-GDP ratios raises the specter of fiscal dominance by tightening the links between fiscal policy, monetary policy and government debt management. When higher policy interest rates or depreciating currencies raise concerns about debt sustainability, monetary independence is compromised. Possible manifestations of these concerns include the 'fear of floating,' fiscal pressure against policy interest rate hikes, financial repression, and the like.

Applying the public finance logic of the second-best, the mitigation of a constraint like the 'Original Sin' revives borrowing opportunities: it furthers the integration of EMs into global financial markets. However, it may also induce secondary effects with more ambiguous welfare effects. Indeed, the resulting increase in the external debt of EMs raises concerns that volatile sovereign spreads and interest rates may drive EMs' fiscal vulnerability. Such fragility is evident in growing susceptibility to confidence crises when a seemingly moderate level of aggregate external debt/GDP ratio may push a country into a foreign debt crisis, i.e., the presence of multiple equilibria. Lower global risk tolerance in the risk-off environments, also known as a flight to quality, deteriorating growth prospects of EMs can sharply widen sovereign spreads and risk premia, inducing capital flight and exacerbating roll over difficulties. These events may put in motion self-fulfilling confidence crises of increasing sovereign spreads, leading to a sudden stop and capital flight crisis within just a few quarters. The end game frequently saddles the public sector with a massive debt overhang associated with bailing out the financial system and prime corporate borrowers, sometimes in the context of IMF stabilization packages [see Aizenman et al. (2019)].

Appendix Figure 1. Foreign and domestic holding of sovereign debt
This figure plots the foreign and domestic holding of central government debt in emerging market from Arslanalp and Poghosyan (2014) in dashed and solid line respectively.

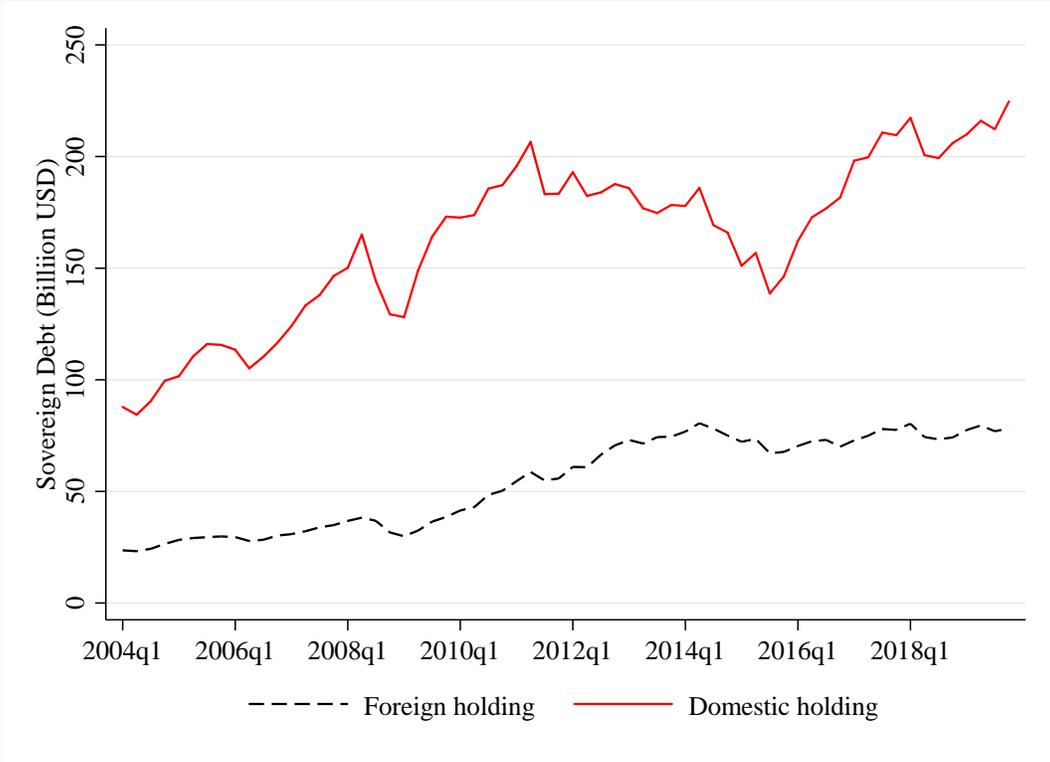

Appendix Figure 2. Sovereign debt denominated in foreign and local currency.
The dashed and solid lines plot respectively the foreign- and local-currency denominated central government debt in emerging market from Arslanalp and Poghosyan (2014).

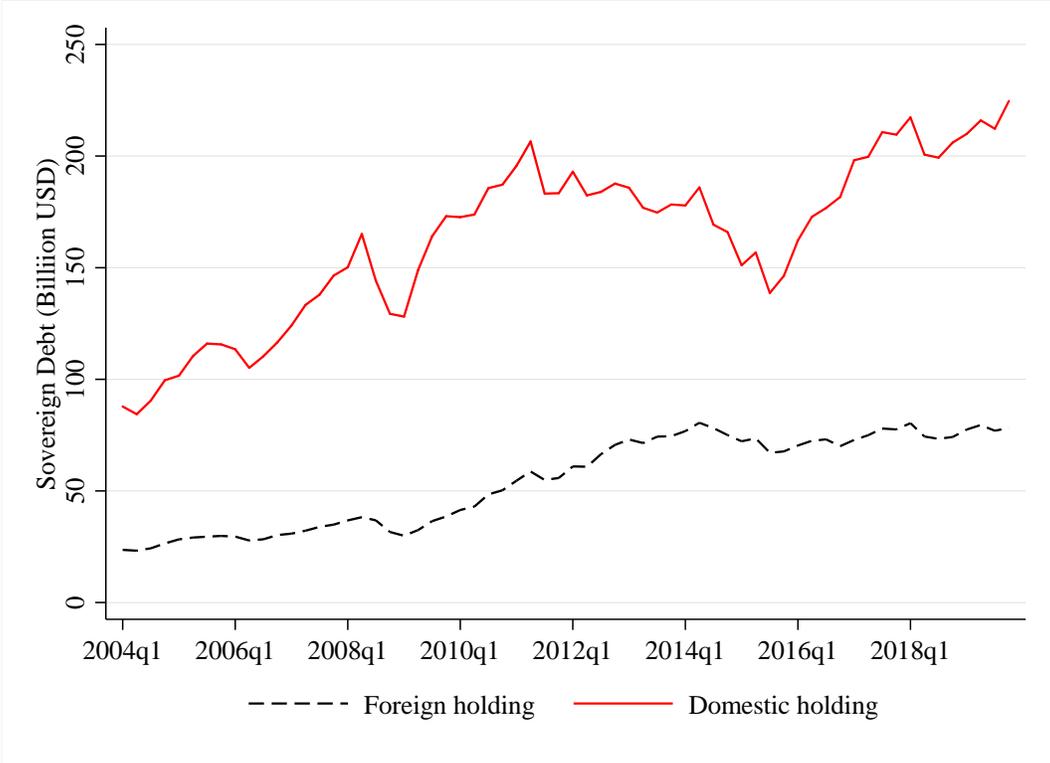

Appendix Figure 1. Foreign and domestic holding of sovereign debt by currency decomposition.

The left (right) panel demonstrates the foreign (domestic) holding of sovereign debt by local and foreign currency.

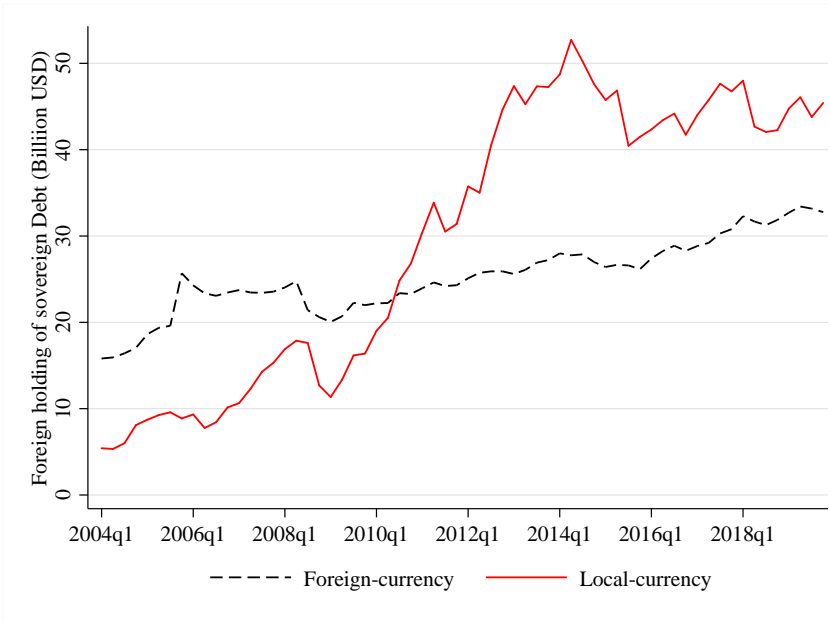
Panel A: Foreign holding of sovereign debt

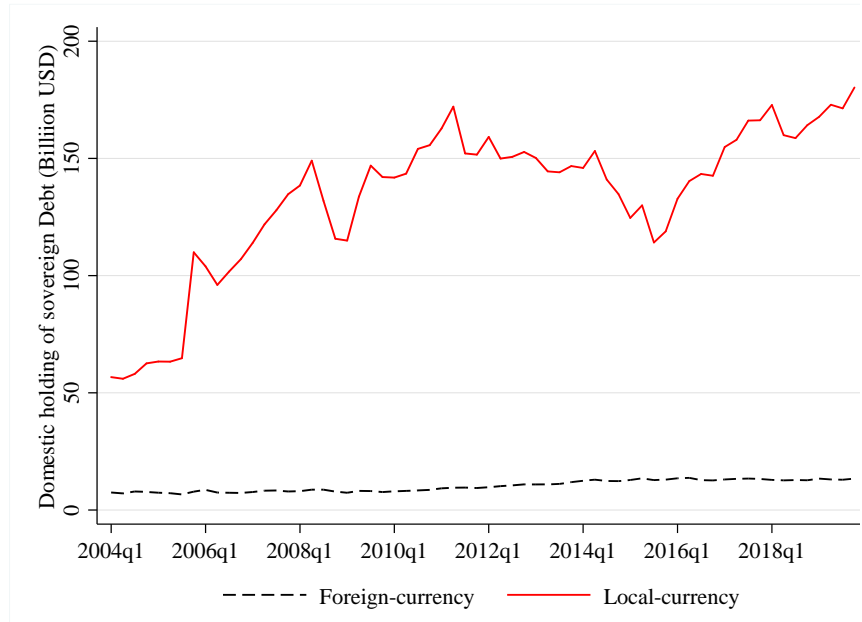
Panel B: Domestic holding of sovereign debt

# Appendix Table 1: Marginal effect Baseline results

This table reports the estimation marginal effects from the following probit regression:

$$P(D_{i,j,t} = 1) = \beta\ X_{j,t} + \gamma S_{i,j,t} + C_j + T_t + \varepsilon_{i,j,t}\ (1),$$

where $D_{i,j,t} = 1$ bond $i$ in country $j$ at period $t$ is issued in local-currency. The key country-specific variable is $X_{j,t}$, which takes the value of (i) $FX_{j,t}$, the currency appreciation of country $j$ at period $t$ relative to USD; (ii) $Yield_{j,t}$ 10-year sovereign bond yield difference between country $j$ and US at period $t$; and (iii) $IT_{j,t}$, a dummy variable that equals 1 if country $j$ is pursuing inflation targeting at period $t$. The vector of bond-level control variable $S_{i,j,t}$ covers (i) log ($Size_{i,j,t}$), the logarithm of the issued amount of bond $i$ in country $j$ at period $t$; (ii) log ($Maturity_{i,j,t}$), the logarithm of the maturity of bond $i$ in country $j$ at period $t$; and (iii) $Zero_{i,j,t}$, a dummy that equals to one if the bond $i$ in country $j$ at period $t$ is a zero-coupon bond. All regression control for $C_j$ and $Y_t$, the country and year fixed effects respectively. Standard errors reported in the parenthesis are clustered by country. ***, ** and * denote significance level at 1%, 5% and 10%.

|  | Dependent variable: | | | |
| --- | --- | --- | --- | --- |
|  | (1) | (2) | (3) | (4) |
| FX | 0.183*** |  |  | 0.185*** |
|  | (0.063) |  |  | (0.063) |
| Yield |  | -0.001 |  |  |
|  |  | (0.001) |  |  |
| IT |  |  | 0.003 | 0.004 |
|  |  |  | (0.005) | (0.005) |
| log(Size) | -0.011*** | -0.005*** | -0.011*** | -0.011*** |
|  | (0.001) | (0.001) | (0.001) | (0.001) |
| log(Maturity) | -0.017*** | -0.009*** | -0.018*** | -0.017*** |
|  | (0.002) | (0.001) | (0.002) | (0.002) |
| Zero | 0.039*** | 0.034*** | 0.043*** | 0.039*** |
|  | (0.004) | (0.004) | (0.004) | (0.004) |

# Appendix Table 2: Controlling for lagged currency valuations.

This table reports the estimation results from the following probit regression:
$$P(D_{i,j,t} = 1) = \beta\ X_{j,t} + \gamma S_{i,j,t} + C_j + T_t + \varepsilon_{i,j,t}$$
where $D_{i,j,t} = 1$ bond $i$ in country $j$ at period $t$ is issued in local-currency. The key country-specific variable is $X_{j,t}$, which takes the value of $FX_{j,t}$, the currency appreciation of country $j$ at period $t$ relative to USD and its one- and two-year lags, L1.FX and L2.FX. The vector of bond-level control variable $S_{i,j,t}$ covers (i) log ($Size_{i,j,t}$), the logarithm of the issued amount of bond $i$ in country $j$ at period $t$; (ii) log ($Maturity_{i,j,t}$), the logarithm of the maturity of bond $i$ in country $j$ at period $t$; and (iii) $Zero_{i,j,t}$, a dummy that equals to one if the bond $i$ in country $j$ at period $t$ is a zero-coupon bond. The variable $C_j$ and $Y_t$ are country and year fixed effects respectively. Chi-squared and the associated p value from the Wald Test on the joint significance of currency appreciation related variables are reported in the bottom row of this table. Standard errors reported in the parenthesis are clustered by country. ***, ** and * denote significance level at 1%, 5% and 10%.

|  | (1) | (2) | (3) | (4) |
|---|---|---|---|---|
| FX |  |  | 1.885 | 1.383 |
|  |  |  | (1.808) | (1.919) |
| L1.FX | 3.883*** | 4.559*** | 3.075** | 3.842** |
|  | (1.126) | (1.546) | (1.372) | (1.850) |
| L2.FX |  | -1.307 |  | -1.111 |
|  |  | (1.440) |  | (1.477) |
| log(Size) | -0.253*** | -0.240*** | -0.254*** | -0.241*** |
|  | (0.016) | (0.017) | (0.016) | (0.017) |
| log(Maturity) | -0.375*** | -0.369*** | -0.377*** | -0.370*** |
|  | (0.038) | (0.038) | (0.038) | (0.038) |
| Zero | 1.054*** | 1.057*** | 1.050*** | 1.055*** |
|  | (0.112) | (0.112) | (0.112) | (0.112) |
|  | 8.691*** | 8.326*** | 8.757*** | 8.376*** |
| Constant | (0.480) | (0.470) | (0.485) | (0.475) |
| Observations | 15,041 | 14,961 | 15,041 | 14,961 |
| Log Likelihood | -1,177.820 | -1,160.718 | -1,177.288 | -1,160.460 |
| Wald test Chi-Squared | 11.894 | 5.590 | 6.508 | 3.920 |
| Wald test p-value | 0.001*** | 0.004*** | 0.001*** | 0.008*** |
| Country fixed effects? | Yes | Yes | Yes | Yes |
| Year fixed effects? | Yes | Yes | Yes | Yes |

**Appendix Table 3: Controlling for fiscal position and government expenditure**

This table reports the estimation results from the probit regression:
$$P(D_{i,j,t} = 1) = \beta \ FX_{j,t} + \gamma S_{i,j,t} + \tau DF_{j,t} + C_j + T_t + \varepsilon_{i,j,t},$$
where $D_{i,j,t} = 1$ bond $i$ in country $j$ at period $t$ is issued in local-currency. $FX_{j,t}$ is the currency appreciation of country $j$ at period $t$ relative to USD. The domestic factor $DF_t$ includes (i) Fiscal surplus, the fiscal position normalized by GDP; (ii) government expenditure, the government spending normalized by GDP; (iii) Investment, the domestic investment normalized by GDP; and (iv) Growth, the GDP per capita growth rate. The vector of bond-level control variable $S_{i,j,t}$ covers (i) log $(Size_{i,j,t})$, the logarithm of the issued amount of bond $i$ in country $j$ at period $t$; (ii) log $(Maturity_{i,j,t})$, the logarithm of the maturity of bond $i$ in country $j$ at period $t$; and (iii) Zero$_{i,j,t}$, a dummy that equals to one if the bond $i$ in country $j$ at period $t$ is a zero-coupon bond. The variable $C_j$ and $Y_t$ are country and year fixed effects respectively. Standard errors reported in the parenthesis are clustered by country. ***, ** and * denote significance level at 1%, 5% and 10%.

| | (1) | (2) | (3) |
|---|---|---|---|
| FX | 7.599** | 6.617** | 7.337** |
| | (3.383) | (3.007) | (3.540) |
| log(Maturity) | -0.200*** | -0.219*** | -0.200*** |
| | (0.021) | (0.019) | (0.021) |
| log(Size) | -0.465*** | -0.366*** | -0.464*** |
| | (0.052) | (0.041) | (0.052) |
| Zero | 0.836*** | 1.014*** | 0.837*** |
| | (0.160) | (0.118) | (0.160) |
| Fiscal surplus | 6.926 | | 6.268 |
| | (7.088) | | (7.556) |
| Government Expenditure | | 0.426 | -0.152 |
| | | (0.481) | (0.581) |
| Growth | -0.037* | -0.012 | -0.035 |
| | (0.021) | (0.017) | (0.022) |
| Investment | 0.006*** | 0.007*** | 0.005*** |
| | (0.002) | (0.002) | (0.002) |
| Constant | 8.063*** | 7.550*** | 8.139*** |
| | (0.886) | (1.017) | (0.931) |
| Observations | 10,860 | 12,904 | 10,860 |
| Log Likelihood | -695.605 | -982.928 | -695.571 |
| Country fixed effects? | Yes | Yes | Yes |
| Year fixed effects? | Yes | Yes | Yes |

**Appendix Table 4: Addressing incident parameter problem with bias correction.**

This table reports the probit regression that corrects for the bias caused by incident parameter problem following Hahn and Newey (2004) based on the following model:

$$P(D_{i,j,t} = 1) = \beta\ X_{j,t} + \gamma S_{i,j,t} + C_j + T_t + \varepsilon_{i,j,t}\ (1),$$

where $D_{i,j,t} = 1$ bond $i$ in country $j$ at period $t$ is issued in local-currency. The key country-specific variable is $X_{j,t}$, which takes the value of (i) $FX_{j,t}$, the currency appreciation of country $j$ at period $t$ relative to USD; (ii) $Yield_{j,t}$ 10-year sovereign bond yield difference between country $j$ and US at period $t$; and (iii) $IT_{j,t}$, a dummy variable that equals 1 if country $j$ is pursuing inflation targeting at period $t$. The vector of bond-level control variable $S_{i,j,t}$ covers (i) log ($Size_{i,j,t}$), the logarithm of the issued amount of bond $i$ in country $j$ at period $t$; (ii) log ($Maturity_{i,j,t}$), the logarithm of the maturity of bond $i$ in country $j$ at period $t$; and (iii) $Zero_{i,j,t}$, a dummy that equals to one if the bond $i$ in country $j$ at period $t$ is a zero-coupon bond. The variable log ($Size_{i,j,t}$) is dropped in the estimation due to computational matrix singularity. All regression control for $C_j$ and $Y_t$, the country and year fixed effects respectively. Standard errors reported in the parenthesis are clustered by country. ***, ** and * denote significance level at 1%, 5% and 10%.

|  | Dependent variable: $P(D_{i,j,t} = 1)$ | | | |
| --- | --- | --- | --- | --- |
|  | (1) | (2) | (3) | (4) |
| FX | 4.156*** |  |  | 4.138*** |
|  | (1.135) |  |  | (1.128) |
| Yield |  | -0.020 |  |  |
|  |  | (0.024) |  |  |
| IT |  |  | 0.182* | 0.179* |
|  |  |  | (0.101) | (0.103) |
| log(Maturity) | -0.388*** | -0.260*** | -0.395*** | -0.389*** |
|  | (0.035) | (0.051) | (0.034) | (0.035) |
| Zero | 1.050*** | 2.273*** | 1.036*** | 1.058*** |
|  | (0.111) | (0.320) | (0.109) | (0.110) |
| Observations | 12,724 | 9,949 | 12,882 | 12,724 |
| Average individual effect | 3.649 | 3.371 | 8.081 | 3.688 |
| Country fixed effects? | Yes | Yes | Yes | Yes |
| Year fixed effects? | Yes | Yes | Yes | Yes |

**Appendix Table 5: Heterogeneity before and after Global Financial Crisis (GFC).**

This table reports the estimation results from the following OLS regression

$$D_{i,j,t} = \beta\ X_{j,t} + \vartheta\ X_{j,t} * GFC_t + \gamma S_{i,j,t} + C_j + T_t + \varepsilon_{i,j,t}, \quad (1)$$

where $D_{i,j,t} = 1$ bond $i$ in country $j$ at period $t$ is issued in local-currency. $GFC_t$ is a dummy variable that equals to 1 after 2007 and 0 otherwise. The country-specific variable is $X_{j,t}$, which takes the value of (i) $FX_{j,t}$, the currency appreciation of country $j$ at period $t$ relative to USD; (ii) $Yield_{j,t}$ 10-year sovereign bond yield difference between country $j$ and US at period $t$; and (iii) $IT_{j,t}$, a dummy variable that equals 1 if country $j$ is pursuing inflation targeting at period $t$. The vector of bond-level control variable $S_{i,j,t}$ covers (i) $\log(Size_{i,j,t})$, the logarithm of the issued amount of bond $i$ in country $j$ at period $t$; (ii) $\log(Maturity_{i,j,t})$, the logarithm of the maturity of bond $i$ in country $j$ at period $t$; and (iii) $Zero_{i,j,t}$, a dummy that equals to one if the bond $i$ in country $j$ at period $t$ is a zero-coupon bond. Wald test and the p-value are from the test of null hypothesis that the sum of FX (yield, IT) and its interaction with GFC is 0. The variable $C_j$ and $Y_t$ are country and year fixed effects respectively. Standard errors reported in the parenthesis are clustered by country. ***, ** and * denote significance level at 1%, 5% and 10%.

|  | (1) | (2) | (3) |
| --- | --- | --- | --- |
| FX | 0.758*** | | |
|  | (0.108) | | |
| Yield | | -0.003*** | |
|  | | (0.001) | |
| IT | | | 0.042*** |
|  | | | (0.008) |
| FX*GFC | -1.357*** | | |
|  | (0.336) | | |
| Yield*GFC | | 0.003** | |
|  | | (0.001) | |
| IT*GFC | | | -0.024*** |
|  | | | (0.008) |
| GFC | 0.125*** | -0.043 | -0.165 |
|  | (0.033) | (0.045) | (0.128) |
| log(Size) | -0.009*** | -0.006*** | -0.010*** |
|  | 0.0004 | (0.0003) | (0.0004) |
| log(Maturity) | -0.018*** | -0.012*** | -0.015*** |
|  | (0.002) | (0.002) | (0.002) |
| Zero | 0.119*** | 0.088*** | 0.119*** |
|  | (0.006) | (0.006) | (0.006) |
| Observations | 15,072 | 12,297 | 15,270 |

| Adj. R-squared        | 0.194       | 0.149           | 0.226       |
|-----------------------|-------------|-----------------|-------------|
| Country fixed effects? | Yes         | Yes             | Yes         |
| Year fixed effects?   | Yes         | Yes             | Yes         |
|                       | FC+FC*GFC   | Yield+Yield*GFC | IT+IT*GFC   |
| Wald test             | 54.541      | 42.488          | 27.910      |
| p-valued              | 0.000***    | 0.000***        | 0.000***    |